\documentclass[conference]{IEEEtran}
\IEEEoverridecommandlockouts
\usepackage{cite}
\usepackage{amsmath,amssymb,amsfonts}
\usepackage{algorithmic}
\usepackage{graphicx}
\usepackage{textcomp}
\usepackage{xcolor}

\usepackage{multirow}    

\def\BibTeX{{\rm B\kern-.05em{\sc i\kern-.025em b}\kern-.08em
    T\kern-.1667em\lower.7ex\hbox{E}\kern-.125emX}}
\begin{document}

\title{Performance Analysis of Quantum Support Vector Classifiers and Quantum Neural Networks
\thanks{
This work has been supported by the State Research Agency of the Spanish Government (Grant PID2023-147422OB-I00) and by the Xunta de Galicia (Grant ED431C 2022/44), supported by the EU European Regional Development Fund (ERDF). DAE has also received support from project RYC2022-038121-I, funded by MCIN/AEI/10.13039/501100011033 and
European Social Fund Plus (ESF+).
CITIC, as a center accredited for excellence within the Galician University System and a member of the CIGUS Network, receives subsidies from the Department of Education, Science, Universities, and Vocational Training of the Xunta de Galicia. Additionally, it is co-financed by the EU through the FEDER Galicia 2021-27 operational program (Ref. ED431G 2023/01). 
}
}

\author{
\IEEEauthorblockN{Tomás Villalba-Ferreiro}
\IEEEauthorblockA{
\textit{Universidade da Coruña (CITIC)}\\
A Coruña, Spain \\
tomas.villalba.ferreiro@udc.es}
\and
\IEEEauthorblockN{Eduardo Mosqueira-Rey}
\IEEEauthorblockA{
\textit{Universidade da Coruña (CITIC)}\\
A Coruña, Spain \\
eduardo@udc.es}
\and
\IEEEauthorblockN{Diego Alvarez-Estevez}
\IEEEauthorblockA{
\textit{Universidade da Coruña (CITIC)}\\
A Coruña, Spain \\
diego.alvareze@udc.es}
}

\maketitle

\begin{abstract}
This study explores the performance of Quantum Support Vector Classifiers (QSVCs) and Quantum Neural Networks (QNNs) in comparison to classical models for machine learning tasks. By evaluating these models on the Iris and MNIST-PCA datasets, we find that quantum models tend to outperform classical approaches as the problem complexity increases. While QSVCs generally provide more consistent results, QNNs exhibit superior performance in higher-complexity tasks due to their increased quantum load. Additionally, we analyze the impact of hyperparameter tuning, showing that feature maps and ansatz configurations significantly influence model accuracy. We also compare the PennyLane and Qiskit frameworks, concluding that Qiskit provides better optimization and efficiency for our implementation. These findings highlight the potential of Quantum Machine Learning (QML) for complex classification problems and provide insights into model selection and optimization strategies.
\end{abstract}


\section{Introduction}

Quantum computing has emerged as a promising field with the potential to revolutionize various computational tasks, including Machine Learning (ML). Classical machine learning models have been widely successful; however, they face limitations when dealing with high-dimensional and complex datasets. Quantum machine learning (QML) \cite{biamonte2017quantum} aims to leverage the unique properties of quantum mechanics, such as superposition, interference and entanglement, to improve model efficiency and performance.

Among QML approaches, Quantum Support Vector Classifiers (QSVCs) and Quantum Neural Networks (QNNs) implemented using Variational Quantum Circuits (VQCs) have gained significant attention. QSVCs utilize quantum kernels to enhance classical support vector machines, whereas QNNs employ parameterized quantum circuits to learn data representations. While both methods incorporate quantum advantages, they differ in their structure, optimization complexity, and computational efficiency. Understanding the strengths and limitations of these models is crucial for determining their applicability to real-world problems.

In this paper, we have developed a performance analysis of QSVCs and QNNs using datasets such as the Iris dataset and MNIST-PCA. These datasets are widely used for demonstration purposes in classical ML, and their reduced dimensions make them ideal for experiments in the current Noise Intermediate-Scale Quantum (NISQ) era. This is because the availability of quantum computers with a large number of reliable qubits is limited today. Simulators can be used instead to overcome these limitations. However, the maximum number of qubits that can be simulated on a classical computer is in the range of 30-50, depending on the entanglement properties of the underlying quantum circuit. This means that it is currently not possible to consider using large datasets to run QML models on them.

We investigate the effectiveness of QSVCs and QNNs compared to classical models in classification tasks to assess whether quantum models provide an advantage as problem complexity grows. Furthermore, we conduct a comparative evaluation of QSVCs and QNNs, examining their accuracy, consistency, and sensitivity to hyperparameter tuning. Our study also explores the impact of different quantum feature maps and ansatz configurations on model performance, shedding light on optimal parameter choices. Additionally, we compare the PennyLane and Qiskit frameworks to determine which is more suited for efficient quantum model training and potential hardware execution. The findings of this research contribute to a better understanding of the role of quantum models in machine learning and provide insights into optimizing their performance. By highlighting the conditions where quantum approaches excel, we aim to support further advancements in quantum-enhanced machine learning techniques.

The paper is structured as follows: Section \ref{sec:qml} details the materials and methods used in this study, that is, the datasets used and a description of the Quantum Support Vector Machines and Quantum Neural Networks used. Section \ref{sec:iris} describes the Iris classification experiment and Section \ref{sec:mnist} the MNIST-PCA classification experiment. We wrap up with Section \ref{sec:discussion} dedicated to discussion and conclusions.

\section{Quantum Machine Learning}
\label{sec:qml}

\subsection{Datasets}

The Iris flower is a multivariate dataset used by the British statistician and biologist Ronald Fisher in his 1936 paper \cite{fisher1936use}, which is widely used in ML as a simple test case for many classification techniques. The dataset consists of 50 samples from each of three Iris species (\textit{Iris setosa}, \textit{Iris virginica}, and \textit{Iris versicolor}). Four features were measured from each sample: the length and width of the sepals and petals, all measured in centimeters.

The goal of using this toy dataset is to confirm that quantum models can solve basic classification problems and compare them to classical ones. For this dataset we expect both approaches to perform well, but for classical to take the lead, as these simple problems are not the forte of quantum computing.

The MNIST-PCA dataset is a simplified version of the famous MNIST dataset \cite{deng2012mnist}, which recently has been used to benchmark QML models \cite{bowles2024better}. This modified version consists of digits 3 and 5 and has a reduced number of features, from 2 to 20, generated by fitting a PCA dimensionality reduction model to the original MNIST training sets.

We use this dataset to check the efficiency of each model in scenarios that more closely resemble real-world problems. Here, we expect the performance of the QML models to be closer to their classical counterparts, as the complexity of the dataset increases.

\subsection{Quantum Neural Networks}

Quantum Neural Networks are implemented in quantum computers by means of Variational Quantum Algorithms (VQAs). A variational algorithm is a near-term algorithm that can be executed on current quantum computers in concert with classical computers. Using VQAs we can build a Variational Quantum Classifier (VQC) that is a quantum machine learning model similar to a simplified classical ANN and that is composed of the following main parts \cite{mitarai2018quantum}:

\begin{itemize}
    \item \textbf{Feature Map}. That encodes classical input data into a quantum state using a quantum feature map. The choice of feature map affects how well the quantum circuit can separate different classes in Hilbert space. 
    \item \textbf{Parameterized Quantum Circuit (Ansatz)}. The ansatz is a trainable quantum circuit that consists of parameterized gates (such as rotation gates) and entangling operations. These parameters are optimized during training to improve classification accuracy.
    \item \textbf{Measurement Layer}. After the quantum state evolves through the ansatz, measurements are performed in the computational basis to extract relevant information for classification. The measurement outcomes are used to compute expectation values, which are then fed into a classical optimizer.
    \item \textbf{Cost Function}. A classical cost function (such as the cross-entropy loss) evaluates the difference between predicted and actual labels. This guides the parameter optimization process.
    \item \textbf{Classical Optimizer}. A classical optimization algorithm (e.g., gradient descent, COBYLA, SPSA, or Adam) is used to iteratively update the variational parameters to minimize the cost function.
\end{itemize}

\subsubsection{Feature map}

Throughout our experiments we are going to use two different feature maps, the ZFeatureMap and the ZZFeatureMap. Both are among the most popular choices in the related literature, easy to understand but complex enough to get good results. Using these feature maps we can also compare the difference between using entanglement (with the ZZFeatureMap) and not using it (with the ZFeatureMap).

\begin{enumerate}
    \item \textbf{ZFeatureMap:} this is a simple feature map that applies Phase (P) and Hadamard (H) gates but does not use entanglement (Figure \ref{fig:ZFeatureMap}). These Phase gates are a generalization of a rotation gate, where the P gate performs a rotation around the Z-axis in the Bloch sphere.
    \begin{figure}[htbp]
      \centering
      \includegraphics[width=0.5\linewidth]{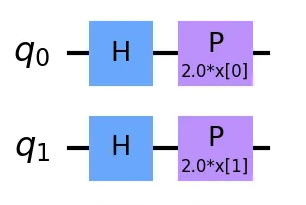}
      \caption{Visualization of a simple ZFeatureMap}
      \label{fig:ZFeatureMap}
    \end{figure}
    \item \textbf{ZZFeatureMap:} this one is a more complex version of the ZFeatureMap that also applies entanglement to the qubits. As we can see in Figure \ref{fig:ZZFeatureMap} apart from the H and P gates ZZFeatureMap used also CNOT gates to add entanglement to the circuit.
    \begin{figure*}[htbp]
      \centering
      \includegraphics[width=\linewidth]{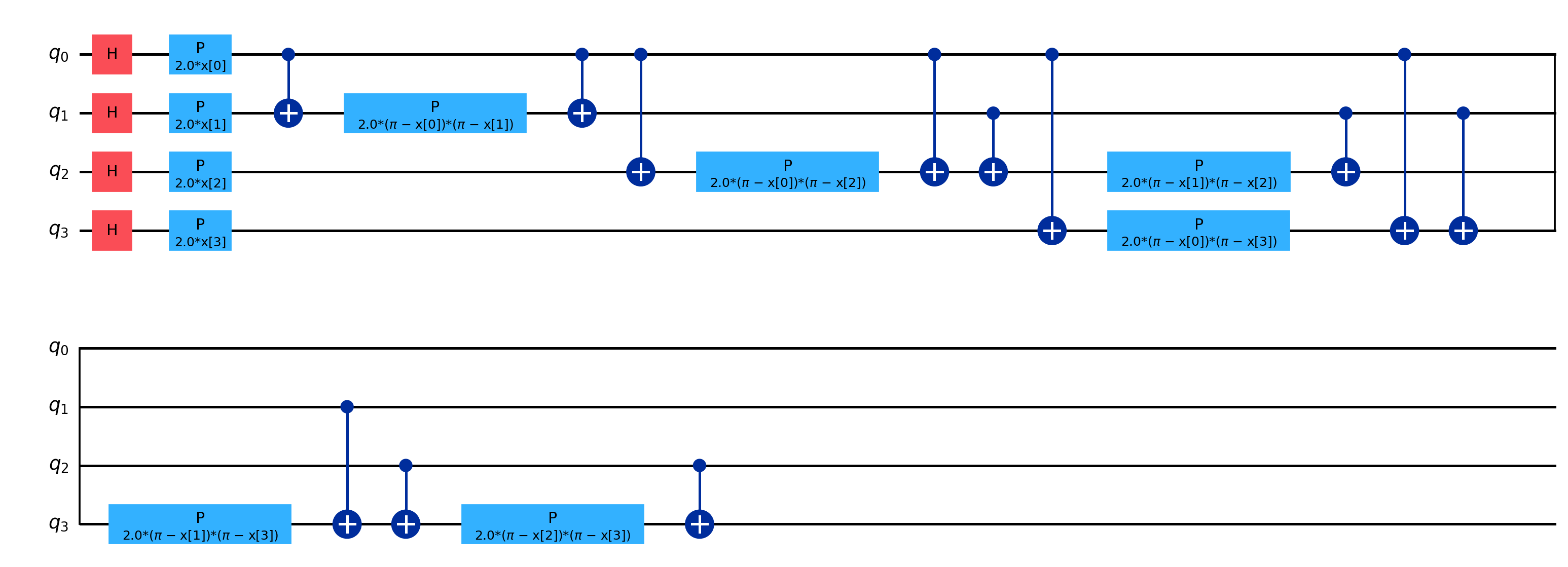}
      \caption{Visualization of a ZZFeatureMap circuit}
      \label{fig:ZZFeatureMap}
    \end{figure*}
\end{enumerate}

\begin{figure}[htbp]
  \centering
  \includegraphics[width=0.75\linewidth]{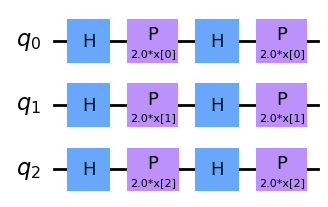}
  \caption{Visualization of a ZFeatureMap circuit with 2 repetitions and 3 features}
  \label{fig:ZFeatureMap_3F_2R}
\end{figure}

These feature maps have 2 hyperparameters, the number of features and the number of layer repetitions. For example, in Figure \ref{fig:ZFeatureMap_3F_2R} we can see a ZFeatureMap for 3 features (the values are depicted as [0], [1] and [2]) and with 2 repetitions.

\subsubsection{ansatz}

The ansatz is also a quantum parametrized circuit, but unlike the feature maps, the ansatz parameters are trainable, i.e., our classifier algorithm will try to obtain the appropriate value for them so that the whole quantum algorithm (feature map and ansatz) works as a classifier for the data supplied to it. We are also going to use the following parametrized circuits in the ansatz:

\begin{enumerate}
    \item \textbf{RealAmplitudes:} Its name comes from the fact that it only works with real numbers, its quantum states have the complex part as 0 at all times. This parametrized circuit is composed of two types of gates, Y rotation gates and CNOT gates; these last ones are the ones that add the entanglement to the circuit. As with the feature map, one can also select the number of repetitions and features of the ansatz. For example, in Figure \ref{fig:real_amplitudes} we can see a RealAmplitudes circuit with 3 repetitions and 4 features.

    \begin{figure}[htbp]
      \centering
      \includegraphics[width=\linewidth]{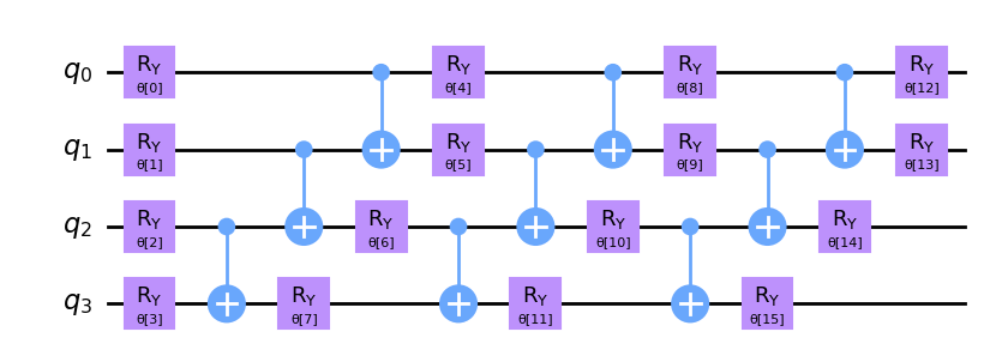}
      \caption{Visualization of a real amplitudes circuit with 3 repetitions and 4 features }
      \label{fig:real_amplitudes}
    \end{figure}

    \item \textbf{EfficientSU2:} this parametrized circuit is normally used for VQCs for classification. 
    This comes from the Special Unitary group of degree 2 (SU2), formed by 2x2 matrices with determinant one, for example the Pauli rotation gates. This circuit is formed by SU2 gates and CNOT gates to add the entanglement to the circuit.
    
    \begin{figure}[htbp]
      \centering
      \includegraphics[width=\linewidth]{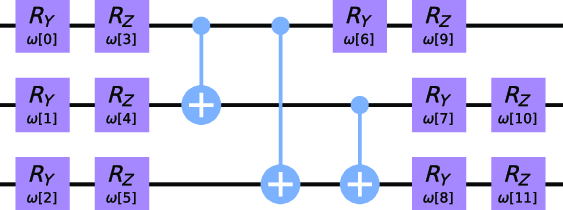}
      \caption{Visualization of a efficient SU2 circuit with 1 repetition and 3 features }
      \label{fig:efficient_SU2}
    \end{figure}
    
    You can select the SU2 gates to use, but in our case we are just going to use the Y and Z rotation gates. In Figure \ref{fig:efficient_SU2}, we can see a EfficientSU2 circuit with 1 repetition and 3 features. 

    \begin{figure}[htbp]
      \centering
      \includegraphics[width=\linewidth]{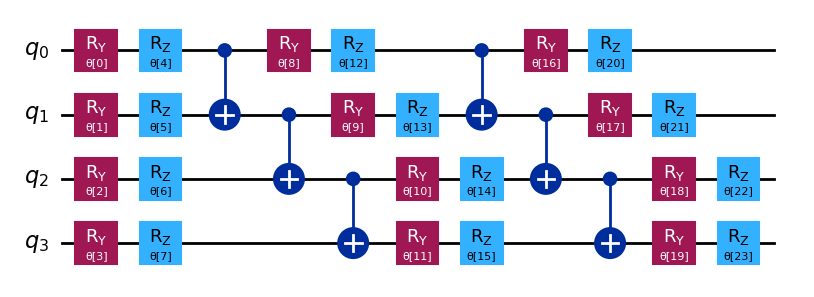}
      \caption{Visualization of a efficient SU2 circuit with 2 repetitions and 4 features }
      \label{fig:efficient_SU2_2R_4F}
    \end{figure}
    
    If we increase the number of repetitions, new instances of the circuit will be added between the two sets of rotation gates. This is illustrated in Figure \ref{fig:efficient_SU2_2R_4F}.   
\end{enumerate}

\subsubsection{Loss function}

As the loss function, we are going to use the \textbf{cross-entropy loss}. This function measures the performance of a classification model whose output is a probability value between 0 and 1. If we have \emph{N} number of classes, and we take \(target_i\) as a binary value (either 0 or 1) that indicates whether the expected result is class \emph{i}, and \(predict_i\) as the predicted probability of the classifier, the formula for the cross-entropy loss looks like this:

\[
L = - \sum_{i=0}^{N_{\text{classes}}} target_i \cdot \log(predict_i)
\]

\subsubsection{Optimizer}

Finally, as optimizer we will use the Simultaneous Stochastic Perturbation Approximation (SPSA) algorithm, as it works well in situations where the objective function is noisy, which is normally the case for variational quantum algorithms.  

\subsection{Quantum Support Vector Machines}

One promising approach in QML is the Quantum Support Vector Machine (QSVM), which extends classical support vector machines (SVMs) by using quantum kernels to compute more efficiently on high-dimensional feature spaces. These quantum kernels can provide computational advantages for certain classification tasks, making QSVMs an attractive candidate for complex pattern recognition problems. Recent studies have explored the effectiveness of quantum kernels in enhancing classical models, demonstrating their potential to outperform traditional approaches under certain conditions \cite{havlivcek2019supervised, schuld2019quantum}.

The idea behind a quantum feature map is to use a quantum circuit to embed the input data in a quantum Hilbert space. Once in this space, one can compute the corresponding inner product between the induced data pairs using specific circuit constructs \cite{havlivcek2019supervised}, effectively implementing a quantum kernel. One can then use the resulting quantum kernel matrix as in classical kernel methods. For example, to train a Support Vector Machine. The main idea is that by using a feature map that is hard to simulate classically, one can obtain an advantage over its classical counterpart.

For the quantum support vector machine models we are going to use two different approaches: 
\begin{enumerate}
    \item \textbf{Regular Quantum Kernel Estimation:} In this approach we follow the Compute-Uncompute approach described in \cite{havlivcek2019supervised} to compute the individual entries of the quantum kernel matrix, and then pass the resulting matrix to a classical SVM solver.

    \item \textbf{Trainable Quantum Kernel:} Finding the appropriate quantum feature mapping for a machine learning task is generally difficult \emph{a priori}. In this approach, we add additional parameters to the quantum kernel to fine-tune it using an iterative algorithm. For this purpose, we follow the procedure described in \cite{Glick_2024}. We then use the resulting quantum kernel to train an SVM, following the same procedure as in the previous case.

    \begin{figure*}[htbp]
      \centering
      \includegraphics[width=\linewidth]{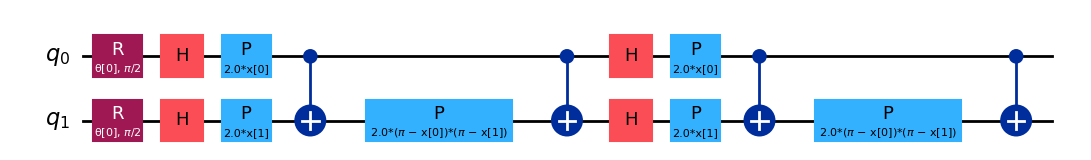}
      \caption{The result of adding two parametrized Ry gates to a ZZFeatureMap with two repetitions and two features}
      \label{fig:log_loss_function}
    \end{figure*}

    More specifically, we use a modification of the ZZFeatureMap that adds two Ry gates as trainable parameters, both of which depend on the same variable. In Figure \ref{fig:log_loss_function} we can see the circuit resulting from the combination of our parameterized circuit with the ZZFeatureMap. As initial point for the optimization algorithm, we set the Ry gates to the value \(\frac{\pi}{2}\). Theoretically, one would hope to get better results with this approach than with a simple feature map.
    
\end{enumerate}

\section{Iris classification experiment}
\label{sec:iris}

The Iris dataset is widely used for simple experiments and demonstrations in ML due to its well-known feature and class distribution. The Setosa class is linearly separable from the other two, making it relatively easy to classify. However, the Versicolor and Virginica classes are slightly non-linearly separable, making their classification more challenging.

For our procedures, we normalize the data using a MinMaxScaler within the range [0,1]. The dataset is split using a 80/20\% partition, with 120 samples for training and 30 for testing. Since the dataset consists of only four features, we increase the number of experiments for each model to 500 in order to obtain more reliable results.

\subsection{Quantum kernels}

\subsubsection{Regular QSVCs}

We use Qiskit's QSVC class with a precomputed quantum kernel with both the Z and ZZ feature maps. We experiment with different number of circuit repetitions for each case, from 1 to 5. The number of qubits will be the same for all of them, as they have to match the number of features in the dataset, in this case four. We performed 500 executions of each experiment, resulting in a total of 2500 executions for both the ZFeatureMap and the ZZFeatureMap.

\subsubsection{Quantum Trained Kernels}

In addition to the normal QSVCs, we also look at the trained quantum kernel. This quantum kernel, as explained in the previous sections, is composed of a parametrized quantum circuit and either a ZFeatureMap or a ZZFeatureMap. For each feature map we similarly use different number of circuit repetitions, from 1 to 5. The number of qubits stays the same as for the regular QSVCs, four. In total, we did 2500 executions for the Z feature map and another 2500 for the ZZ. We used the SPSA optimizer with a learning rate of 0.01 and a maximum of 10 iterations.

\subsection{Quantum neural networks}

There are two key considerations when working with Quantum Neural Networks (QNNs). First, their training times are significantly longer than those of QSVCs. As previously discussed, VQCs operate similarly to Artificial Neural Networks (ANNs), where the ansatz is repeatedly executed in a loop until a stopping criterion is met. This results in multiple executions of the parameterized quantum circuit.

Second, VQCs introduce an additional element—the ansatz—which adds more hyperparameters, including the type of ansatz and the number of repetitions. Due to these factors, each combination of feature map, ansatz, and their respective repetitions will be executed 10 times to ensure reliable results.

\subsection{Results}

Due to space limitations, we will not detail all results obtained for all combinations of experiments. Instead, we will detail below which models obtained the best results within each category, and which combination of hyperparameters was used to obtain the results: 

\begin{enumerate}
    \item \textbf{Support Vector Classifier (SVC)}: with a RBF kernel, classical baseline model to compare with quantum kernel methods. Note that in all cases we use the default $C=1$.
    \item \textbf{Support Vector Classifier (QSVC)}: Using ZFeatureMap with two repetitions.
    \item \textbf{QSVC with trained kernel (QSVCTK)}: Using ZFeatureMap with two repetitions.
    \item \textbf{Artificial Neural Network (ANN)}: Two hidden layers with ten neurons each.
    \item \textbf{Variational Quantum Classifier (VQC)}: Using ZFeatureMap and EfficientSU2 with four repetitions each. 
\end{enumerate}

Considering these models in Table \ref{tab:best-iris} we can see a summary of the results obtained by each of them for the specified metrics:

\begin{table}[htbp]
\setlength{\tabcolsep}{3pt}
\centering
\renewcommand{\arraystretch}{1.5}
\begin{tabular}{|l|c|c|c|c|c|}
\hline
\multicolumn{6}{|c|}{\textbf{Classical vs Quantum}} \\ \hline
 & \multicolumn{2}{|c|}{\textbf{Classical models}} & \multicolumn{3}{|c|}{\textbf{Quantum models}}\\ \hline
\textbf{Model} & \textbf{ANN} & \textbf{SVC} & \textbf{QSVC} & \textbf{QSVCTK} & \textbf{VQC} \\
\hline
\textbf{Accuracy}                & 0.9521 & 0.9598 & 0.9589 & 0.9546 & 0.9233 \\ \hline
\textbf{Failure Rate}            & 0.0479 & 0.0402 & 0.0411 & 0.0454 & 0.0767 \\ \hline
\textbf{Sensitivity (Recall)}    & 0.9534 & 0.9612 & 0.9608 & 0.9561 & 0.9317 \\ \hline
\textbf{Specificity}             & 0.9765 & 0.9807 & 0.9802 & 0.9780 & 0.9638 \\ \hline
\textbf{PPV}                     & 0.9572 & 0.9630 & 0.9622 & 0.9584 & 0.9331 \\ \hline
\textbf{NPV}                     & 0.9777 & 0.9811 & 0.9804 & 0.9785 & 0.9652 \\ \hline
\textbf{F1-Score}                & 0.9494 & 0.9572 & 0.9568 & 0.9521 & 0.9200 \\ \hline
\end{tabular}
\caption{Macro Average Metrics for the best hyperparameters combination of each model for the Iris classification}
\label{tab:best-iris}
\end{table}

From Table \ref{tab:best-iris}, we observe that the SVC achieves the highest accuracy, though all models perform similarly, with differences of less than 0.01—except for the VQC. The VQC is the least effective, achieving an accuracy of 92.33\%. Overall, for the Iris classification task, classical and quantum models perform comparably. However, the more quantum-intensive model, the VQC, delivers the lowest accuracy, whereas the QSVC—a hybrid approach where only the kernel is quantum—emerges as the best quantum model, closely following the classical SVC, with a difference of less than 0.10\%.

\section{MNIST-PCA}
\label{sec:mnist}

The original MNIST dataset is more complex than Iris. The MNIST-PCA variant that we use in our experiments is a reduced version consisting of two classes (the digits 3 and 5) with a reduced number of features obtained by applying PCA to the original MNIST dataset. We have tried several models with different numbers of resulting features, from 2 to 20, and found out that we have got the best results when using 5. Therefore, we will use the dataset with 5 features for all executions and we normalize them using a MinMaxScaler within the range [0,1].

\subsection{Quantum kernels}

Because of the higher complexity of this dataset, in comparison to Iris, leading to some of the models to take much longer training times, we will here execute each model only once using the predefined train-test split, which consists of 11,552 training elements and 1,902 test elements, with an equal class balance.

\subsection{Quantum neural networks}

With the QNNs (implemented as VQCs), we followed the same methodology as with the Iris dataset. One aspect to highlight among all the related experiments is that the best results were always obtained with ansatzes that had a high number of repetitions. This result suggests that the higher the number of parameters for the trainable part of the circuit, the higher the learning capacity of the models.

\subsection{Results}

Similar to the Iris dataset, it is not possible to detail the full results obtained for all the classical and the quantum model experiments. Therefore, we will detail which models obtained the best results within each category and indicate which combinations of hyperparameters were used to obtain the results, which for the MNIST-PCA dataset were as follows: 

\begin{enumerate}
    \item \textbf{Support Vector Classifier (SVC)}: with a RBF kernel.
    \item \textbf{Support Vector Classifier (QSVC)}: ZFeatureMap with 3 repetitions.
    \item \textbf{QSVC with trained kernel (QSVCTK)}: ZZfeatureMap with 1 repetition.
    \item \textbf{Artificial Neural Network (ANN)}: Two hidden layers with 49 and 12 neurons, respectively.
    \item \textbf{Variational Quantum Classifier (VQC)}: Using ZFeatureMap with 3 repetitions and EfficientSU2 anstaz with 10 repetitions.
\end{enumerate}

Considering these models in Table \ref{tab:best-mnist} we can see a summary of the results obtained by each of them for the specified metrics:

\begin{table}[htbp]
\setlength{\tabcolsep}{3pt}
\centering
\renewcommand{\arraystretch}{1.5}
\begin{tabular}{|l|c|c|c|c|c|}
\hline
\multicolumn{6}{|c|}{\textbf{Classical vs Quantum}} \\ \hline
 & \multicolumn{2}{|c|}{\textbf{Classical models}} & \multicolumn{3}{|c|}{\textbf{Quantum models}}\\ \hline
\textbf{Model} & \textbf{ANN} & \textbf{SVC} & \textbf{QSVC} & \textbf{QSVCTK} & \textbf{VQC} \\
\hline
\textbf{Accuracy}                & 0.9099 & 0.9132 & 0.9264 & 0.9227 & 0.9338   \\ \hline
\textbf{Failure Rate}            & 0.0901 & 0.0868 & 0.0736 & 0.0773 & 0.0662   \\ \hline
\textbf{Sensitivity (Recall)}    & 0.9829 & 0.9837 & 0.9832 & 0.9776 & 0.9518   \\ \hline
\textbf{Specificity}             & 0.8455 & 0.8510 & 0.8762 & 0.8743 & 0.9178   \\ \hline
\textbf{PPV}                     & 0.8493 & 0.8546 & 0.8752 & 0.8729 & 0.9109   \\ \hline
\textbf{NPV}                     & 0.9825 & 0.9835 & 0.9833 & 0.9779 & 0.9557   \\ \hline
\textbf{F1-Score}                & 0.9111 & 0.9144 & 0.9261 & 0.9223 & 0.9309   \\ \hline
\end{tabular}
\caption{Macro Average Metrics for the best hyperparameters combination of each model for the MNIST-PCA classification}
\label{tab:best-mnist}
\end{table}

\section{Discussion and conclusions}
\label{sec:discussion}

Our findings indicate that QSVCs and VQCs can achieve performance comparable to, if not better than, the classical models tested.

Comparing the results of the Iris and MNIST-PCA datasets, we observe that quantum models become more advantageous as the complexity of the problem increases. For the relatively simpler Iris dataset, the difference in accuracy between classical and quantum models is minimal —less than 0.01. However, for the more complex MNIST-PCA dataset, the quantum models outperform the classical models by a margin of up to 0.020.

Among quantum models, QSVCs generally perform better and exhibit greater consistency than VQCs. The latter require tuning of a larger number of hyperparameters, including both the feature map and the ansatz. This added complexity makes VQCs more difficult to optimize, and even after this optimization, they often perform similarly to QSVCs.

Despite their lower overall performance, VQCs become more competitive as problem complexity increases. For the Iris dataset, QSVCs outperform VQCs in accuracy by 0.033, whereas for the MNIST-PCA dataset, VQCs improve QSVCs by 0.074. This trend mirrors the broader comparison between quantum and classical models, reinforcing the idea that increasing the quantum load enhances performance in more complex scenarios. 
VQCs, which consist of a feature map and an ansatz, can be considered to involve a higher quantum component (except for the classical loss and optimizer) than QSVCs, which leverage a quantum kernel but otherwise rely on a classical SVM for training. Our results suggest that as the quantum component of a model increases, whether from classical to quantum models, or from QSVCs to VQCs, performance improves as the problem complexity increases.

Regarding the hyperparameter optimization, the tuning of hyperparameters is crucial for both QSVCs and VQCs. Our analysis shows that when the ZFeatureMap is used, the optimal number of repetitions is two —exceeding this number degrades performance. With the ZZFeatureMap, a single repetition yields the best results, although a local maximum often appears around four repetitions. Beyond this point, performance declines. Notably, when using a trained kernel, the model sensitivity to feature map repetitions decreases, but the optimal range remains between one and five. In general, the ZFeatureMap outperforms the ZZFeatureMap in all of our experiments. This result, which seems to question the possibility of quantum advantage related to entanglement, has also been observed in the recent work of \cite{bowles2024better}. 

Unlike the feature maps, ansatz performance improves with an increasing number of repetitions. This pattern is especially evident in the MNIST-PCA dataset, where the best results are obtained with ten repetitions in the ansatz.

Finally, we have made a framework comparison between PennyLane and Qiskit. We do not include many details of this comparison throughout the paper due to space limitations, but in general we can state that our results suggest that Qiskit's implementation offers better performance than Pennylane's. Qiskit's \texttt{FidelityStatevectorKernel} is highly optimized for quantum kernel applications with the statevector simulator, significantly accelerating training times compared to PennyLane's standard \texttt{FidelityQuantumKernel}. Additionally, Qiskit is better suited for execution on actual quantum hardware, facilitating future research that may involve deploying our models on quantum computers. PennyLane would have been a strong candidate if we had intended to integrate with AI libraries such as TensorFlow. However, since this was not a requirement for our study, Qiskit emerged in this case as the best choice.

To summarize the final conclusions of our work, we can state that: 

\begin{itemize}
    \item Quantum models perform as well as, or better, than classical models, with increasing advantages as the problem complexity grows.
    \item QSVCs are more consistent and excel in simpler tasks, whereas VQCs perform better on complex datasets like MNIST-PCA.
    \item Increasing the quantum load improves performance: VQCs, with an arguably higher quantum component than QSVCs, benefit more from challenging classification problems.
    \item Hyperparameter tuning is crucial: for creating a feature map, lower number of repetitions are required, but for the ansatz performance improves with higher number of repetitions.
    \item Qiskit is the preferred framework, since it offers faster training and better optimization for quantum hardware execution. PennyLane could be useful for more general AI-library integration.
\end{itemize}

For future research, several directions could be explored to extend the findings of this study:

\begin{itemize}
    \item \textbf{Evaluation on larger and more complex datasets}: Expanding the analysis to higher-dimensional datasets to assess the scalability and effectiveness of quantum models.
    \item \textbf{Optimization of variational circuits}: Investigating new ansatz architectures and optimization techniques to enhance VQC performance and training stability.
    \item \textbf{Hybrid quantum-classical approaches}: Exploring different methods to integrate quantum and classical machine learning, such as quantum feature extraction combined with deep learning models.
    \item \textbf{Execution on real quantum hardware}: Testing the models on actual quantum processors to evaluate their practical viability and resilience to quantum noise.
    \item \textbf{Benchmarking with alternative quantum kernels}: Assessing different quantum kernel designs and their impact on QSVC performance.
    \item \textbf{Integration with AI frameworks}: Studying the benefits of incorporating PennyLane with TensorFlow or PyTorch for hybrid quantum-classical learning pipelines.
    \item \textbf{Resource efficiency analysis}: Investigating the trade-off between computational cost, training time, and model accuracy to optimize quantum machine learning workflows.
\end{itemize}

\bibliographystyle{IEEEtran}
\bibliography{2025-IJCNN-Roma-QNNs}

\end{document}